\documentclass[11pt]{article}
\usepackage[utf8]{inputenc}
\usepackage{amssymb,amsmath}
\usepackage{bm} %math bold symbols
\usepackage{booktabs} %some alignment stuff
\usepackage{array}
\usepackage{latexsym}
\usepackage{graphicx}
\usepackage{color}
\usepackage{datetime}
\usepackage[nosort]{cite}
\usepackage{verbatim}
\usepackage{enumerate}
\usepackage{chngpage} % allows for temporary adjustment of side margins
\usepackage{mathrsfs}
\usepackage{euscript}
\usepackage{psfrag}
\usepackage{soul}

\usepackage{cite}
\usepackage{tikz}

\usepackage{datetime}

\usepackage[nosort]{cite}
\usepackage{chngpage} % allows for temporary adjustment of side margins
\usepackage{setspace}
\usepackage{tensor}
\usepackage{physics}

\usepackage{mciteplus}

\usepackage[colorlinks=true,      linkcolor=blue,      urlcolor=blue,      
            filecolor=blue,      citecolor=blue,       pdfstartview=FitH,     
						pdfpagemode=UseNone,      bookmarksopen=true]{hyperref}  % LINKS
\usepackage[all]{hypcap}     %should be loaded AFTER hyperref, which should otherwise be loaded last.

\definecolor{cardinal}{rgb}{0.6,0,0}
\definecolor{darkgreen}{rgb}{0,0.4,0}
\definecolor{golden}{rgb}{0.92, 0.7, 0}
\definecolor{midnight}{rgb}{0, 0, 0.5}
\definecolor{darkblue}{rgb}{0, 0, 0.7}
\definecolor{purple}{rgb}{0.5, 0, 0.5}

\def\IR{\mathbb{R}}

\def\n_1{{p}}

\numberwithin{equation}{section}

\begin{document}

\phantom{AAA}
\vspace{-10mm}

\begin{flushright}
%
%IPHT-T19/021\\
%
\end{flushright}

\vspace{1.9cm}

\begin{center}

{\huge {\bf Positive Spin-induced Quadrupole Moment}}

\medskip

{\huge {\bf in String Theory}}

{\huge {\bf \vspace*{.25cm}  }}

\vspace{1cm}

{\large{\bf { Iosif Bena and Ang\`ele Lochet}}}
 
\vspace{1cm}

Institut de Physique Th\'eorique, \\
Universit\'e Paris-Saclay, CEA, CNRS,\\
Orme des Merisiers, Gif sur Yvette, 91191 CEDEX, France \\[12pt]

\vspace{10mm} 
{\footnotesize\upshape\ttfamily iosif.bena @ ipht.fr,  angele.lochet @ ipht.fr,}    \\

\vspace{2.2cm}
 
\textsc{Abstract}
\end{center}

\noindent

\begin{adjustwidth}{3mm}{3mm} % to adjust the L and R margins

\vspace{-1.2mm}
\noindent

We identify singularity-free Running-Kerr-Taub-Bolt solutions of eleven-dimensional supergravity that descend to four-dimensional rotating solutions with flat-space asymptotics. We compute their spin-induced quadrupole moment and find that for a certain range of charges this quadrupole moment is positive. 
This behavior differs from the Kerr black hole and from most other spinning objects constructed with ``normal'' four-dimensional matter, and we discuss the top-down physics of these solutions that could be responsible for this unusual behavior. 

\end{adjustwidth}

%\end{titlepage}
\thispagestyle{empty}
\newpage

%%%%%%%%%%%%%%%%%%%%%%%%%%%%%%%%%%%%%

\tableofcontents

\section{Introduction}

Gravitational waves observed by LIGO/VIRGO/KAGRA confirm the overall features of General Relativity. However, since the observed mergers happen in the rotation plane, the constraints on the spin-induced quadrupole moment of the merging objects are rather weak. Intriguingly, they appear to favor a {\em positive} quadrupole moment \cite{LIGOScientific:2021sio}, which has the opposite sign of that of Kerr black holes, $M_2 = - J^2 /M$. 
 
Such a positive quadrupole moment is rather unusual, since objects made of ``normal'' matter (such as neutron stars) tend to pancake when spinning, and hence have a negative quadrupole moment. In contrast, an object with a positive quadrupole moment becomes thinner when spinning. Hence the ``matter'' that makes it has rather unusual properties. 

It is intriguing to ask whether String Theory can provide such matter. Indeed, in String Theory one can construct many ``black hole microstate geometries'' that have the same charges, mass and angular momenta as black holes, but no horizon (see \cite{Bena:2022rna} for a recent review). These ten-dimensional solutions have nontrivial topology and fluxes, which support them against immediate collapse into a black hole. And from a four-dimensional perspective, the higher-dimensional topology and fluxes appear as very unusual matter: for example, some centers of the smooth eleven-dimensional supersymmetric microstates correspond, when reduced to four dimensions, to D6 branes of negative mass and negative charge
\cite{Bena:2005va,Berglund:2005vb,Bena:2007kg}. In supersymmetric solutions these branes are in equilibrium because their gravitational {\em repulsion} compensates their electric attraction.  This unusual matter allows the microstate geometries to avoid collapsing into a black hole \cite{Bena:2013dka}. 

The main result of this paper is to show that spinning this unusual matter can give rise to solutions with a positive spin-induced quadrupole moment.

There is a simple illustration of this phenomenon in the real world: in imponderability one can place a ball of liquid in equilibrium inside a denser liquid. If one spins this system around the axis of the ball of light liquid, this ball becomes thinner, thus acquiring a positive spin-induced quadrupole moment\footnote{We thank Denis Vion for this example.}.

The spinning solutions we construct are surrounded by vacuum. By analogy with the spinning ball of liquid, the positivity of their spin-induced quadrupole moment indicates that they contain ingredients that are somehow ``lighter'' than the vacuum. And the intuition coming from supersymmetric solutions confirms this: Objects with negative mass and charge are indeed lighter than the vacuum, and spinning them reduces the gravitational repulsion between various components, makes them thinner. 

To see this behavior explicitly one must construct non-extremal spinning microstate geometries, which is not easy. The largest families of microstate geometries with four-dimensional black-hole asymptotics correspond to supersymmetric black holes  
\cite{Bena:2006kb, 
Bena:2007qc, 
Bena:2010gg,
Bianchi:2016bgx, 
Bianchi:2017bxl, 
Heidmann:2017cxt, 
Bena:2017fvm, 
Avila:2017pwi,
Tyukov:2018ypq,
Warner:2019jll,
Rawash:2022sum} 
and to extremal non-supersymmetric black holes 
\cite{Goldstein:2008fq,
Bena:2009ev,
Bena:2009en,
DallAgata:2010srl,
Vasilakis:2011ki,
Heidmann:2018mtx}, which have infinite AdS$_2$ throats and do not resemble real-world black holes\footnote{The multipole moments of supersymmetric and almost-BPS black holes have been computed in \cite{Bena:2020see,Bianchi:2020bxa,Bena:2020uup,Bianchi:2020miz,Ganchev:2022vrv} and in \cite{Bah:2021jno}, and can be both positive and negative.}.

There has recently been a flourish of activity in constructing systematically microstate geometries for non-extremal black holes 
\cite{Bah:2022pdn,
Heidmann:2021cms,
Bah:2021owp,
Bah:2021rki,
Heidmann:2022zyd,
Bah:2022yji,
Bah:2023ows,
Heidmann:2023kry}, 
but the methods that allow the construction of these solutions do not allow one to easily add rotation in four dimensions. Hence, these solutions have no spin and therefore one cannot talk about their spin-induced quadrupole moment.

To find rotating non-extremal geometries in four dimensions we turn to an older, less known construction: one can build microstate geometries by solving the linear system of BPS or almost-BPS equations on top of any arbitrary Ricci-flat four-dimensional Euclidean space.  These solutions are supersymmetric only when this space is also hyper-K\"ahler \cite{Gauntlett:2002nw, Gutowski:2004yv,Bena:2004de}. 
However, when the four-dimensional base is Euclidean Schwarzschild or Euclidean-Kerr-Taub-Bolt, this gives rise to smooth non-supersymmetric ``Running-Bolt'' solutions \cite{Bena:2009qv}, which can have nontrivial angular momentum when reduced to four dimensions. 

The magnetic fluxes of the Running-Bolt solutions are wrapping the bolt, and are self-dual on the four-dimensional base. Hence, these solutions belong to a different class of solutions than the topological stars constructed in \cite{Bah:2020ogh}. Because of this self-duality, the electric and magnetic fields give rise to a nontrivial velocity along the Kaluza-Klein direction, which only vanishes for some special values of the magnetic fluxes\footnote{This running is not problematic from a four-dimensional perspective, as it can be undone by a boost \cite{Bena:2009qv, Elvang:2005sa}. However, this makes the identification of the charges more subtle than for solutions which do not run.}.

It is important to note that not all smooth rotating five-dimensional Running-Kerr-Taub-Bolt solutions reduce to four-dimensional solutions with black-hole asymptotics. The smoothness conditions in five dimensions imply a nontrivial periodicity relation between the compact Kaluza-Klein angle and the four-dimensional angle along which the solution rotates. Unless a certain relation between the parameters of the Kerr-Taub-Bolt base space is satisfied, the four-dimensional solution will have conical asymptotics.\footnote{Or, if one chooses instead to compactify along a combination of compact and non-compact angles, Melvin-space asymptotics. Note that the recent solutions of \cite{Bianchi:2025uis}  are in this category.} 

The relation between the parameters of the Euclidean-Kerr-Taub-Bolt base space needed to render the four-dimensional solution asymptotically flat can only be satisfied when the signature of this base space  changes from $(+,+,+,+)$ to $(-,-,-,-)$, but this need not scare us: such pathological base-spaces can give rise to smooth geometries in five-dimensions, both for supersymmetric solutions \cite{Bena:2005va,Berglund:2005vb,Bena:2007kg}, and also for Running-Bolt solutions \cite{Bena:2009qv}. Furthermore, it is precisely the presence of regions of $(-,-,-,-)$ signature in the base space that indicates the presence in 3+1 dimensions of objects with negative D6 brane mass and charges, responsible for the positivity of the spin-induced quadrupole moment.

In this paper we  identify the rotating Running-Kerr-Taub-Bolt solutions that have $\IR^{3,1}$  asymptotics when reduced to four dimensions, and compute their spin-induced quadrupole moment. We find that there is an infinite possibility of combinations of the parameters of the Euclidean-Kerr-Taub-Bolt base space that give non-conical asymptotics and are free of closed timelike curves.  

Moreover, we show that all Running-Kerr-Taub-Bolt solutions have positive spin-induced quadrupole moments when the running of the bolt vanishes or is small. As the running intensifies, the sign of the spin-induced quadrupole moment becomes negative (similar to that of the Kerr black hole), and then diverges in the singular limit when the bolt runs with the speed of light.

In Section \ref{section2}  we review the properties of the Kerr-Taub-Bolt Euclidean base spaces that we use to construct Running-Kerr-Taub-Bolt solutions, and we give the metric and fluxes of these solutions. In Section \ref{section3} we compute the charges, angular momenta, and higher multipole moments. In Section \ref{section4} we identify the ranges of fluxes where the solutions are regular, and show that the spin-induced quadrupole moments of these solutions is positive at zero or small running. In Section \ref{section5} we present the supersymmetric limit of Running-Kerr-Taub-Bolt solutions, when they reduce to a Taub-NUT center with smooth fluxes (corresponding in four dimensions to a sixteen-supercharge fluxed D6 brane \cite{Balasubramanian:2006gi}). We conclude in Section \ref{Conclusions}. 

{\bf Note:} In parallel to our work, Heidmann, Pani and Santos are submitting a paper in which they construct new rotating four-dimensional solutions by applying duality transformations on the ``Euclidean-Kerr-Taub-Bolt times time'' geometry  \cite{HPS}. These solutions cannot be obtained from ours by duality transformations, but share with them several properties, in particular the ubiquity of regions of parameters where the spin-induced quadrupole moment, $M_2$, is positive.

\section{The Kerr-Taub-Bolt base space}
\label{section2}
The four-dimensional Euclidean-Kerr-Taub-Bolt (EKTB) solution is Ricci flat, and can be used as a base space for building five-dimensional solutions. Its metric is 
\begin{equation}
\label{EKTB base space}
    ds_{EKTB}^2 = \Xi (\frac{dr^2}{\Delta} + d\theta^2) + \frac{\sin^2\theta}{\Xi} (\alpha d\tau + P_r d\phi)^2 + \frac{\Delta}{\Xi}(d\tau + P_\theta d\phi)^2 ,
\end{equation}
with: 
\begin{align*}
        \Xi &\equiv r^2- (N + \alpha \cos\theta)^2, & \Delta &\equiv r^2-2mr-\alpha^2+N^2,\\
        P_r &\equiv r^2-\alpha^2-\frac{N^4}{N^2-\alpha^2}\ , & P_\theta &\equiv -\alpha \sin \theta ^2 + 2 N \cos \theta -\frac{\alpha N^2}{N^2-\alpha^2}\ .
\end{align*}
The parameters $m$, $\alpha$ and $N$ correspond to the mass, angular momentum and NUT charge of the Lorentzian solution obtained by adding an unwarped time coordinate to this solution and reducing the resulting $4+1$-dimensional solutions to $3+1$ dimensions along $\tau$. The EKTB solution reduces to the Taub-NUT solution when $m=N$ and $\alpha=0$ and to the Euclidean Kerr metric when $N=0$.

The quantization of the NUT charge, or equivalently the absence of non-removable Dirac strings requires (see for details \cite{Gibbons:1979nf,Bena:2009qv}): 
\begin{equation}
\label{periodicity_cond_inf}
    (\tau, \phi)\sim (\tau - 8N\pi, \phi + 2\pi)\, .
\end{equation}
The solution can also have conical singularities when $r$ approaches $r_{+}\equiv m+\sqrt{m^2-N^2+\alpha^2}$, the largest root of the equation $\Delta=0$. By changing coordinates one can show that these singularities are absent when  the periodicity of $\tau$ and $\phi$ is \cite{Gibbons:1979nf,Bena:2009qv}: 
\begin{equation}
\label{periodicity_cond_rplus}
    \left(\tau, \phi\right) \sim \Big( \tau - \frac{2 \pi}{\kappa},\phi + \frac{2 \pi \alpha }{\kappa \abs{P_{r_{+}}}}\Big)\, ,
\end{equation}
with $\kappa \equiv \abs{ \frac{r_{+}-r_{-}}{2 P_{r_{+}}}}$. 

The compatibility of the periodicity lattices~\eqref{periodicity_cond_inf} and \eqref{periodicity_cond_rplus} and the requirement that $\phi$ be $2 \pi$ periodic imply that:
\begin{equation}
\label{periodicity of phi}
 \frac{ \alpha}{\kappa \abs{P_{r_{+}}}} = \frac{ \alpha}{\sqrt{m^2 - N^2 +\alpha^2}}\equiv \n_1 \in {\mathbb Z}
\end{equation}
and 
\begin{equation}
\label{N and kappa}
    \frac{1}{\kappa} = 4  \abs{N}\, .
\end{equation}
%
%{\Golden I added the absolute values because $\kappa$ is positive by def ; RHS needs to be positive}
We can choose without loss of generality $\alpha$ and $\n_1$ to be positive, by changing the orientation of $\tau$ and flipping the sign of $N$. 
We thus have two equations relating $N,m$ and $\alpha$. Since $r_{+}$ is the root of a quadratic equation, there are four branches of solutions \cite{Gibbons:1979nf,Bena:2009qv}. When $N$ and $P_{r_{+}}$ have the same sign, the regular solutions must satisfy:
\begin{multline}
\label{same-sign}
    16 N (N^2-\alpha^2)^2m^3 -4(5N^6-8\alpha^2 N^4+2\alpha^4N^2+\alpha^6)m^2 \\
    - 16 N (N^2-\alpha^2)^3m + 20 N^8-52 \alpha^2 N^6 + 49 \alpha^4 N^4-16 \alpha^6 N^2 = 0
\end{multline}
while when $N$ and $P_{r_{+}}$ have opposite signs the regular solutions must satisfy
\begin{multline}
\label{opp-sign}
    -16 N (N^2-\alpha^2)^2m^3-4(5N^6-8\alpha^2 N^4+2\alpha^4N^2+\alpha^6)m^2 \\
    + 16 N (N^2-\alpha^2)^3m + 20 N^8-52 \alpha^2 N^6 + 49 \alpha^4 N^4-16\alpha^6 N^2 = 0  \, .
\end{multline}
Since equations~\eqref{same-sign}-\eqref{opp-sign} are obtained by squaring the original regularity constraint~\eqref{N and kappa}, there is an extra inequality coming from requiring that the two squared terms have the same sign. Since these equations are homogeneous of degree 8 in $m,N,\alpha$, their solutions can be plotted in units where one takes $N=1$ or $N=-1$. 

\begin{figure*}[h!]
    \centering
    \begin{minipage}[t]{0.49\textwidth}
        \centering
        \includegraphics[width=\linewidth,trim={0.5cm 0.3cm 0 0},clip]{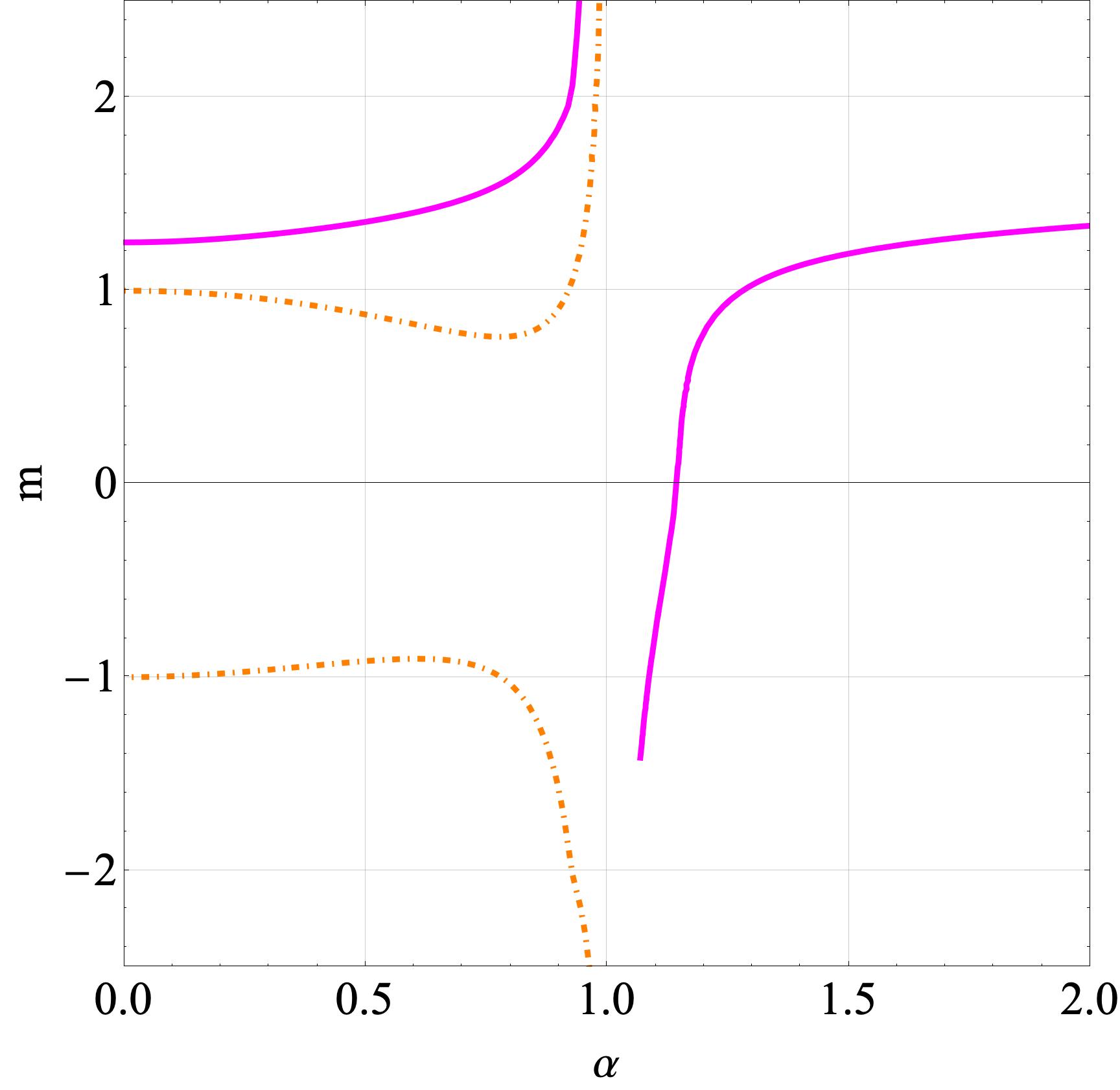}
        \caption{Solutions $m(\alpha)$ to equation (\ref{N and kappa}), for $N$ and $P_{r_{+}}$ having the same sign. Magenta curves have $N=1>0$ and orange dashed curves have $N=-1<0$.}
        \label{fig:N>0 pr+>0 m(alpha)}
    \end{minipage}
    \hfill
    \begin{minipage}[t]{0.49\textwidth}
        \centering
        \includegraphics[width=\linewidth,trim={0.5cm 0.3cm 0 0},clip]{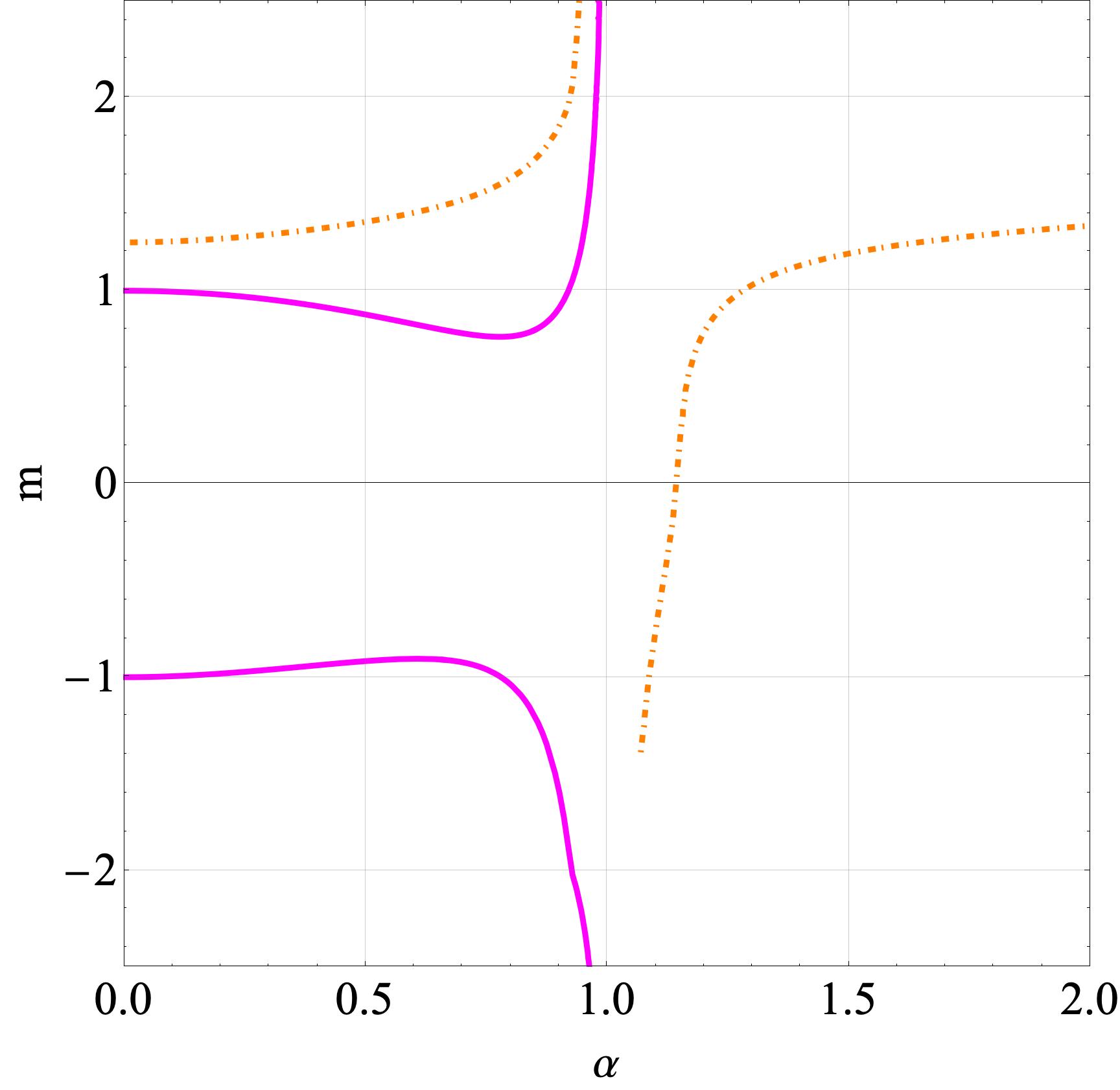}
        \caption{Solutions $m(\alpha)$ to equation (\ref{N and kappa}), for $N$ and $P_{r_{+}}$ having opposite signs. Magenta curves have $N=1>0$ and orange dashed curves have $N=-1<0$.}
        \label{fig:N>0 pr+<0 m(alpha)}
    \end{minipage}
\end{figure*}

There are three possible branches when $N$ and $P_{r_{+}}$ have the same sign, represented on Figure~\ref{fig:N>0 pr+>0 m(alpha)}
and three when $N$ and $P_{r_{+}}$ have opposite signs, represented on Figure~\ref{fig:N>0 pr+<0 m(alpha)}.

\begin{figure*}[h!]

        \centering
        \includegraphics[width=0.6\linewidth,trim={0.5cm 0.3cm 0 0},clip]{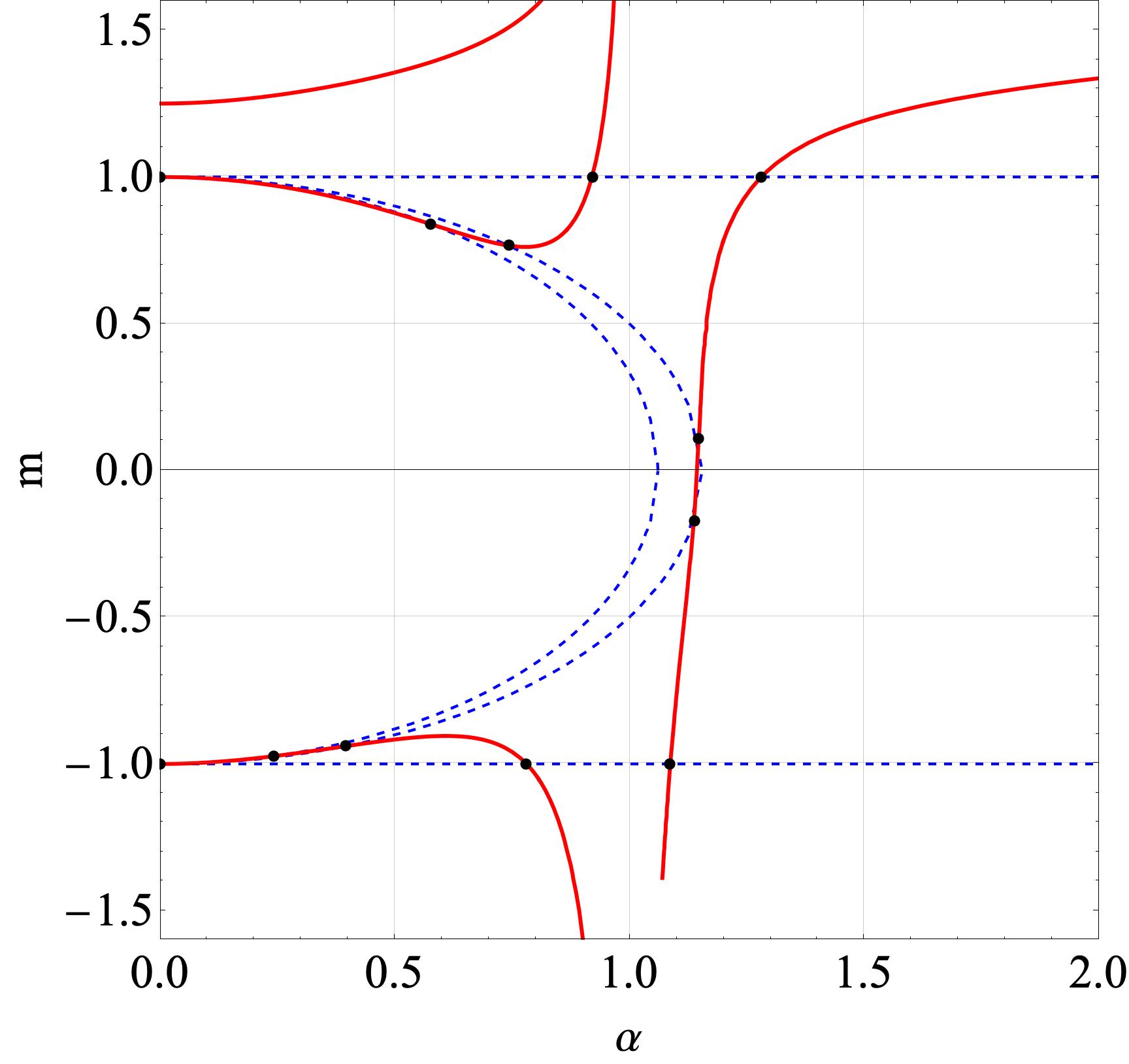}
        \caption{We intersect the $m(\alpha)$ curves from the previous figures, with the condition \eqref{periodicity of phi} represented in  in dashed blue lines. We only plot the curves corresponding to $ \n_1=1, 2,3$ and to $\alpha\geq 0$. The results are independent of the sign of $N$, since the red curve can be obtained either by superposing the $N=1$ magenta curves in Figures \ref{fig:N>0 pr+>0 m(alpha)} and \ref{fig:N>0 pr+<0 m(alpha)}, or by superposing the  $N=-1$ orange curves. The symmetry comes  because $N\rightarrow-N$ and $m\rightarrow-m$ flip  \eqref{same-sign} with \eqref{opp-sign}.}
        \label{fig:points m(alpha) for N>0}

\end{figure*}

Note also that we impose equation \eqref{periodicity of phi}, which corresponds to the blue profiles on Figure~\ref{fig:points m(alpha) for N>0}. In this Section we only study profiles with $\n_1=1,2,3$ for more clarity. We will discuss the results for larger $\n_1$ in Section \ref{sec:more solutions}. 

As we mentioned in the Introduction, the Kerr-Taub-Bolt base spaces satisfying \eqref{periodicity of phi} have necessarily $|N|\geq m$, and hence they will always have regions where the signature changes from $(+,+,+,+)$ to $(-,-,-,-)$. Hence, in the absence of fluxes, the five-dimensional solution obtained by adding a time coordinate to these solutions is singular. However, when fluxes are present, the warp factors and rotation parameter blow up in exactly the precise way to give rise to a regular five-dimensional solution \cite{Bena:2009qv}. We will see in section \ref{ctc} that in the limit when the fluxes are small, this pathology reappears.

As we will show in Section \ref{section5} the solutions with $\abs{\n_1}=1$ have $m=N$ or $m=-N$, and are the same as the supersymmetric Taub-NUT solution. This is not obvious when $\alpha \neq 0$.\footnote{We thank Pierre Heidmann for pointing this out to us.}.

\subsection{Adding fluxes: the equations}

To build Running-Bolt solutions one needs to add self-dual or anti-self-dual fluxes on the Euclidean-Kerr-Taub-Bolt four-dimensional base. 

These fluxes source warp factors, as well as rotation, giving rise to the ten-dimensional solution:
\begin{equation}
ds^2 = -(Z_1 Z_2 Z_3)^{-2/3} (dt + k)^2 + (Z_1 Z_2 Z_3)^{1/3} ds^2_{EKTB} + (Z_1 Z_2 Z_3)^{1/3}\left({ds^2_{T^2_1} \over Z_1}+{ds^2_{T^2_2} \over Z_2}+ {ds^2_{T^2_3} \over Z_3} \right)
\end{equation}
where the three two-tori are wrapped by the magnetic fluxes proportional to $\Theta^{(I)}$.
The fluxes, warp factors $Z_I$ and rotation parameter, $k$, satisfy:\cite{Bena:2004de,Goldstein:2008fq}
\begin{equation}
     \left\{
    \begin{array}{ll}
   \Theta^{I} = \varepsilon *_4 \Theta^{I} \\[1em]
    \hat\nabla^2 Z_I = \frac{1}{2} \varepsilon \,C_{IJK} *_4[\Theta^J \wedge \Theta^K] \\[1em]
   \varepsilon \, dk + *_4 dk =  Z_1 \Theta^1+Z_2 \Theta^2+Z_3 \Theta^3 
    \end{array}
\right. \, ,
\label{eqs fluxes} 
\end{equation}
where $\varepsilon=1$ when the fluxes are self-dual and $\varepsilon=-1$ when they are anti-self-dual.

The self-dual and anti-self dual fields are related by \cite{Bena:2009qv}:
\begin{equation}
    \begin{split}
        F_+\left(N,\theta\right) = F_-\left(-N,\pi+\theta\right)\, .
    \end{split}
\end{equation}
Therefore, one can easily obtain the anti-self-dual solutions from the self-dual one by performing the transformation ($N\rightarrow-N ; \theta \rightarrow\pi+\theta$).

\subsection{Self-dual solutions}

When $\varepsilon=1$ the fluxes, warp-factors and rotation vector of the Running-Kerr-Taub-Bolt solution are \cite{Bena:2009qv}:
\begin{equation}
\label{selfdualsol}
     \left\{
    \begin{array}{ll}
        \Theta^{I} &= \dfrac{q_I}{(r - (N+\alpha \cos \theta))^2}\Big[dr\wedge \left(d\tau+ P_{\theta} d\phi \right) + \sin \theta d\theta \wedge (\alpha d\tau + P_r d\phi)\Big] \\[1em]
        Z_I &= 1 +\dfrac{1}{2}C_{IJK}\dfrac{q_J q_K}{(N- m)\big(r- (N+\alpha \cos \theta)\big)} \\[1em]
        k &= \mu (d\tau + P_\theta d\phi) +\nu d\phi
    \end{array}
\right. \, .
\end{equation}
The scalars entering the angular momentum vector are $\mu$ and $\nu$, given by: 
\begin{equation}
     \left\{
    \begin{array}{ll}
        \mu\equiv &\gamma \Big(1-\dfrac{2N}{r+N+\alpha \cos \theta}\Big)-(q_1+q_2+q_3)\dfrac{r}{\Xi}\\[1em]
        &+~ \dfrac{q_1 q_2 q_3}{2(N-m)^2}\Big[\dfrac{m-N-2\alpha \cos\theta}{\Xi}-\dfrac{2(N-m)}{(r-(N+\alpha \cos \theta))^2}\Big]\\[1em]
        \nu\equiv &\gamma \alpha \sin^2 \theta -\dfrac{\alpha q_1 q_2 q_3 \sin^2\theta}{(N-m)^2(r-(N+\alpha \cos \theta))}\\[1em]
        \gamma\equiv& -\dfrac{q_1+q_2+q_3}{2(N-m)}+\dfrac{q_1 q_2 q_3}{4(N-m)^3}\Big(2-\dfrac{m+N}{r_{+}}\Big)\\
    \end{array}
\right. \, .
\end{equation}

As we will show in Section \ref{section5}, when $m=N$ the Kerr-Taub-Bolt space becomes  Taub-NUT, and the divergence of the solution above illustrates the impossibility of adding self-dual fluxes to anti-Taub-NUT spaces (whose complex structures are self-dual) \cite{Bena:2009ev}.

The self-dual magnetic fluxes of the solution, proportional to $q_I$, have a nontrivial integral along the bolt and any one of the three two-tori, and correspond to M5 brane fluxes in eleven dimensions. Because of the $F \wedge F\wedge A$ term in the supergravity Lagrangian, these self-dual magnetic fluxes source electric charges, whose value is encoded in warp factors $Z_I$. In particular, the M2 brane charges arise exclusively from the presence of the self-dual M5 fluxes, and the $Z_I$ have no singular sources. The rotation parameter, $k$, asymptotes to $\gamma$ when $r \rightarrow \infty$. Therefore $\gamma$ is the rotation velocity of the bolt relative to an observer at rest at infinity. To have sub-luminal rotation, we require $\abs{\gamma}<1$, thus imposing constraints on the charge parameters, $q_I$.

The full five dimensional metric is:
\begin{equation}
\label{five dim metric unboosted}
    ds_5^2 = -Z^{-2}(dt+k)^2+Z V^{-1}(d\tau +P_\theta' d\phi)^2 + Z V ds^2_3
\end{equation}
with $Z \equiv (Z_{1} Z_{2} Z_{3})^{1/3}$, and:
\begin{align*}
        ds^2_3 &= \dfrac{\Delta_\theta}{\Delta}dr^2+ \Delta_\theta d\theta^2+\Delta \sin^2\theta d\phi^2 , & \Delta_\theta &\equiv \Delta + \alpha^2\sin^2\theta\ ,\\
        P_\theta' &= P_\theta + \alpha \dfrac{\Xi}{\Delta_\theta}\sin^2\theta\ , &
        V &\equiv \dfrac{\Xi}{\Delta_\theta}\ .
\end{align*}
As noted in \cite{Bena:2009qv} this metric is not asymptotically at rest. One needs to un-boost it along $\tau$  in order to obtain an asymptotically static frame: 
\begin{equation}
     \left\{
    \begin{array}{ll}
    \hat{\tau} = (1 - \gamma^2)^{1/2} (\tau - \dfrac{\gamma}{1-\gamma^2} t)\\
    \hat{t} = (1 - \gamma^2)^{-1/2} t 
    \end{array}
\right. \,\ \  .
\end{equation}
This is a well-defined coordinate transformation if the bolt has a sub-luminal rotation ($\abs{\gamma}<1$). Then, the five dimensional metric can be written in a form ready for a Kaluza-Klein reduction along $\hat{\tau}$: 
\begin{equation}
    ds^2_5 = \frac{\hat{I}_4}{(Z V)^2}\Big(d\hat{\tau} + \hat{P}_{\theta}d\phi+ \gamma d\hat{t} -\frac{\mu V^2}{\hat{I}_4}(d\hat{t} +\hat{\nu}d\phi)\Big)^2+\frac{V Z}{\hat{I}_4^{1/2}}ds^2_E 
\end{equation}
where the four-dimensional Einstein-frame metric is
\begin{equation}
    ds^2_E = -\hat{I}_4^{-1/2}(d\hat{t}+\hat{\nu}d\phi)^2+\hat{I}_4^{1/2}\big(\frac{\Delta_\theta}{\Delta}dr^2+ \Delta_\theta d\theta^2+\Delta \sin^2\theta d\phi^2\big) \, , 
\end{equation}
and we used: 
\begin{align*}
        \hat{I}_4&= (1-\gamma^2)^{-1}(Z_{}^3 V- \mu^2 V^2)\, , \\  \hat{P}_{\theta} &= (1-\gamma^2)^{1/2} P_\theta'\ , \\ \hat{\nu}&=(1-\gamma^2)^{-1/2} \big(\nu -\alpha \frac{\Xi}{\Delta_\theta}\sin^2\theta \mu\big) \,
        .
\end{align*}
One must ensure that the metric remains positive definite and that no closed timelike curves are present. These consistency conditions impose nontrivial constraints on the allowed values of the fluxes, $q_I$, on the sign of $N$ and on the sign of $m$ and eliminate certain pairs $(m,\alpha)$, as summarized in Table~\ref{tab:results N=1 selfdual}. For self-dual solutions, the absence of closed timelike curves requires that $N$ be positive \cite{Bena:2009qv}. Without loss of generality, one may therefore set $N=1$ and evaluate the multipole moments for the admissible pairs $(m',\alpha')$. 
As we will show, the absence of CTC's also requires that $m$ be positive.

\section{The charges and the multipole moments. }
\label{section3}

The mass, angular momentum and the higher multipole moments of an asymptotically four-dimensional solution can be evaluated either using the Geroch-Hansen method \cite{Geroch:1970cd,Hansen:1974zz}, or using the Thorne method \cite{RevModPhys.52.299}. The two methods are equivalent \cite{Gursel:1983nkl}.

We will use the Thorne method:
If the four-dimensional spacetime is a stationary, axisymmetric space-time, with the right change of variables it is possible to obtain at infinity a specific coordinate system called ``Asymptotically Cartesian to order N'' (AC-N). This system can be shifted to obtain an ``Asymptotically Cartesian and Mass Centered to order N'' (ACMC-N) system of coordinates. The mass-centered-ness of the ACMC-N coordinate system implies that the first mass multipole moment, $M_1$, vanishes. The asymptotic form of this coordinate system is very precise, and can be found for example in \cite{Bena:2020uup}. Once we have this coordinate system, we can read the asymptotic mass and gravitational multipoles directly from the expansion of the metric at infinity. Let us denote by $M_n$ the ``true" multipole moments, measured in an ACMC-N system of coordinates, and by $\tilde M_n$ the ``fake'' multipole moments measured in an AC-N system of coordinates. 

By expanding the four dimensional metric when $r \rightarrow \infty$, we notice that our choice of coordinate system is not AC-N. Guided by the examples in \cite{Bena:2020uup,Heidmann:2025yzd,Bah:2022yji} to go from non AC-N coordinates to AC-N or ACMC-N, we look for new coordinates $(r_s,\theta_s)$ such that:
\begin{equation}
    \left\{
    \begin{array}{ll}
    r_s \sin \theta_s = \sqrt{r^2-\alpha^2}\sin \theta \\
    r_s \cos \theta_s -b = r \cos \theta
    \end{array}
    \right. \, \ .
\end{equation}
Then, one can check by expanding every term of the metric, that this coordinate system is actually AC-$\infty$. 

The constant $b$ parametrizes how much the center of mass is displaced from the center of the coordinate system. In cylindrical coordinates it is simply a shift in the origin of $z$.
To obtain an ACMC-$\infty$ coordinate system, we have to fix this shift by demanding that $M_1$ vanishes.

We are interested by the expansion of the $g_{\hat{t}\hat{t}} $ and $g_{\hat{t}\phi} $ terms to read off the mass, angular momenta and quadrupole moment, $M_2$. The higher quadrupole moments can also be straightforwardly obtained. In the generic AC-$\infty$ coordinates we have: 
\begin{equation}
    \begin{split}
        g_{\hat{t}\hat{t}} &= -1+\frac{2G_4M}{r_s}+ \frac{2\left(G_4\tilde M_1 \cos\theta_s+c_1\right)}{r_s^2}+ \frac{(3 \cos^2{\theta_s}-1)G_4\tilde{M_2}+ c_2 \cos{\theta_s}+c_3}{r_s^3} +O\left(\frac{1}{r_s^4}\right)\\
        g_{\hat{t}\phi} &= -2 \sin^2\theta_s \frac{G_4 S_1}{r_s} + O\left(\frac{1}{r_s^2}\right)
    \end{split}
\end{equation}
where $\tilde M_1$ and $\tilde M_2$ are the ``fake'' multipole moments obtained from an AC-$\infty$  coordinate system which is not mass-centered, and $c_i$ are constants. 

One can find the value of $b$ which mass-centers this coordinate system, by requiring  $\tilde M_1$ to vanish:
\begin{equation}
    b = \frac{ \alpha \left( (N-m) \left(2 N (N-m)- (q_1q_2+q_2q_3+q_1q_3)\right)-2 \gamma  q_1q_2q_3\right)}{(N-m) \left(2 (N-m) \left(2 \gamma ^2 (N-m)+m+ \gamma  (q_1+q_2+q_3)\right)+q_1q_2+q_2q_3+q_1q_3\right)} \ .
\end{equation}
One then obtains:
\begin{align}
    G_4 M &= \frac{1}{4(1-\gamma^2)}\Big(2m +\frac{q_1q_2+q_1q_3+q_2q_3}{N-m}+ 2 \gamma(q_1+q_2+q_3) +4 \gamma^2 (N-m)\Big)\, , \label{M self dual}\\[1em]
    G_4 S_1 &= G_4 J=-\frac{1}{(1-\gamma^2)^{1/2}}\frac{\alpha q_1 q_2 q_3(m+N)}{4(N-m)^2r_{+}}\ ,\\[1em]
    \begin{split}
        G_4 M_2 &= \frac{1}{4 \left(1-\gamma ^2\right)(N-m)^3}\alpha ^2 \Big((N-m)^2 \big((q_1q_2+q_2q_3+q_1q_3)\\[0.8 em]
    &+2 (N-m) \left(2 \gamma ^2 (N-m)+m+ \gamma (q_1+q_2+q_3) \right)\big)\\[0.8 em]
    &-\frac{\left((N-m) \left(-(q_1q_2+q_2q_3+q_1q_3)+2 N (N-m)\right)-2 \gamma  q_1 q_2 q_3\right)^2}{(q_1q_2+q_2q_3+q_1q_3)+2 (N-m) \left(2 \gamma ^2 (N-m)+m+ \gamma (q_1+q_2+q_3) \right)}\Big)\, .
    \end{split}
    \label{M J M2 self dual}
\end{align}

The link between the four and five dimensional constants is: 
\begin{equation}
    G_4 = \frac{G_5}{\sqrt{1-\gamma^2} (8 \pi |N|)} \ .
\end{equation}
Our aim is to evaluate the value of $G_4 M_2$ in smooth regular solutions with four-dimensional asymptotics, and to determine its sign. The quadrupole moment is a function of $q_I,N,m$ and $\alpha$, but is homogeneous of degree three in $|N|$. We introduce new homogeneous parameters
\begin{align*}
        m' &\equiv \frac{m}{\abs{N}}\ ,  & \alpha' &\equiv \frac{\alpha}{\abs{N}}\ , &
        N' &\equiv \frac{N}{\abs{N}}\ , & q' &\equiv \frac{q}{\abs{N}}\ .
\end{align*}
and then we can write: 
\begin{equation}
    G_4 M_2 (m,\alpha,N,q_I) = \abs{N}^3 G_4 M_2 \big(m',\alpha',N' ,q_I'\big)\, .
\end{equation}

\section{Regular solutions}
\label{section4}

For a given choice of base-space parameters the Running-Kerr-Taub-Bolt solutions are determined entirely by their magnetic fluxes, $q_I$. It is easiest to understand the physics of these solutions by setting all these fluxes equal,  $q_I'=q$, and plotting $G_4 M_2$ as a function of $q$ for each pair of $(m',\alpha')$ that gives solutions free of closed timelike curves. 

\subsection{Closed timelike curves}
\label{ctc}

\begin{table}[h!]
    \centering
    \begin{tabular}{|c|c|c|c|c|c|}
        \hline
        \textbf{$m'$} & \textbf{$\alpha'$} & \textbf{CTC} & \textbf{range of $q'$} & \textbf{$M_2$} & \textbf{$R\equiv{M_2 M}/{J^2}$} \\
        \hline
        0.108378& 1.14789& No& $3.2<q'<3.8$&$-\infty<M_2<12.62$ & $-\infty<R<0.18$\\
        \hline
        -0.173648&1.13716 & Yes & & & \\
        \hline
        0.766044& 0.7422&No & $0.76<q'<0.92$& $-\infty<M_2<1.36$&$-\infty<R<0.33$ \\
        \hline
        0.838852&0.5773 & No& $0.79<q'<0.9$& $-\infty<M_2<1.19$&$-\infty<R<0.12$\\
        \hline
        -0.939693& 0.394931& Yes & & & \\
        \hline
       -0.97386 & 0.240929& Yes& & &\\
        \hline
        
    \end{tabular}
    \caption{Presence of closed timelike curves (CTC) and the range of fluxes and quadrupole multipoles for $N=1$ solutions with $\n_1=1,2,3$. The range of allowed $q'$ is fixed by demanding absence of CTCs.}
    \label{tab:results N=1 selfdual}
\end{table}

One finds that the spin-induced quadrupole moment, $M_2$, is positive within a finite interval of $q$, which always includes the value of $q$ for which the solution does not run ($\gamma=0$).
As one increases the running speed, $\gamma$, the quadrupole moment $M_2$ flips sign and becomes negative. As the speed of the running bolt approaches the speed of light ($\gamma=\pm 1$), the solution develops closed timelike curves, and its parameters diverge.

\begin{figure}[h]
    \centering
\includegraphics[width=0.5\linewidth]{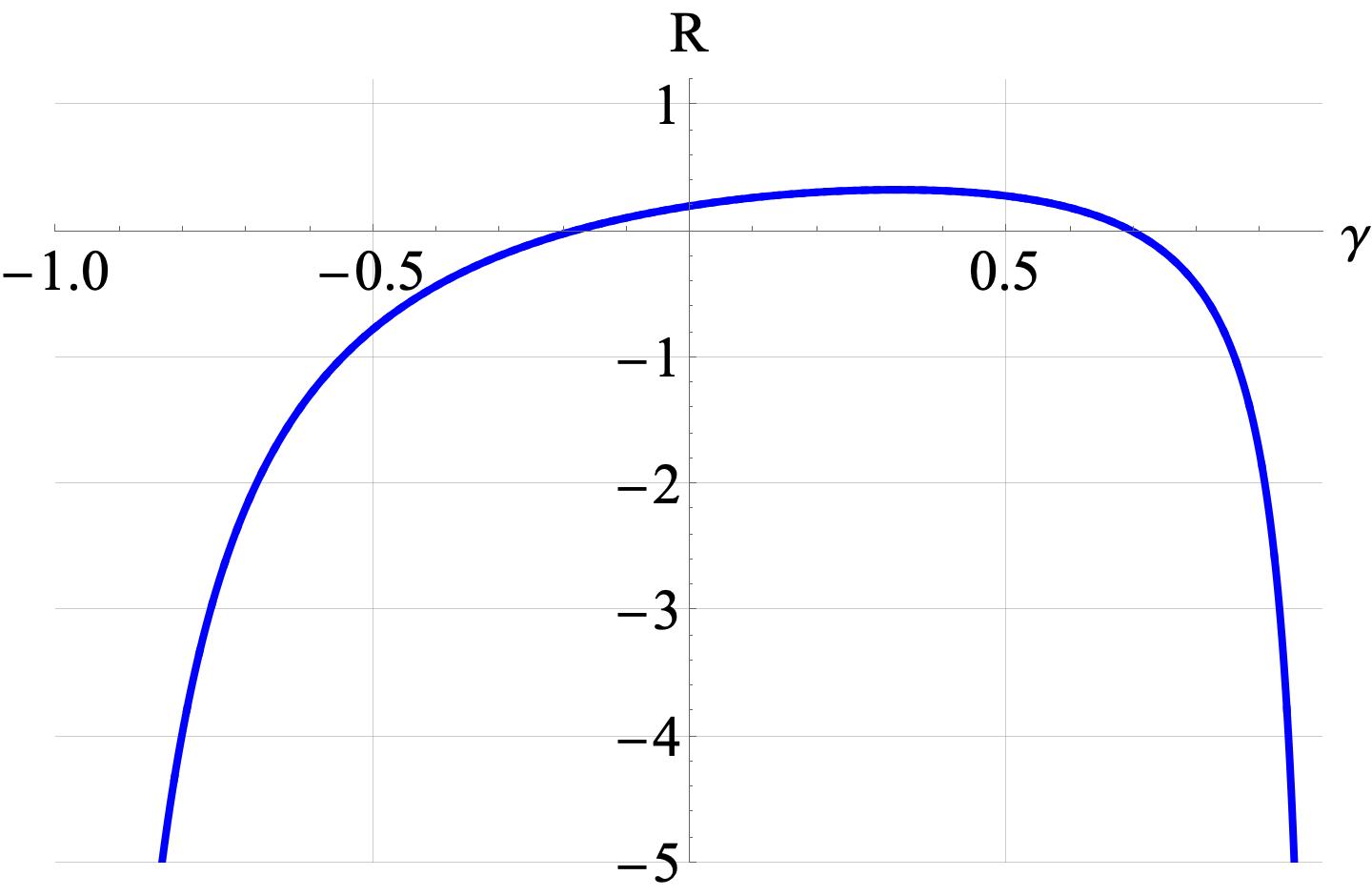}
    \caption{Plot of $R$ with respect to $\gamma$, for the solutions with $(m',\alpha')=(0.766044,0.742227)$ ($\n_1=2$) for $|{\gamma}|< 1$. The ratio diverges when the boost becomes infinite, $|{\gamma}|\rightarrow 1$.}
\label{fig:multipole ratio R (M2)}
\end{figure}

An illustration of this behavior is shown in Figure~\ref{fig:multipole ratio R (M2)}, and the range of $q'$ where the solutions with $\n_1=1,2,3$ are regular is presented in Table~\ref{tab:results N=1 selfdual}.
It is important to emphasize that the solutions excluded by the CTC condition are precisely those with negative mass parameter, $m$. Since the asymptotic mass measured at infinity depends linearly on $+m$, this result is fully consistent with the expectation that the total mass of our solutions be positive. 

As expected, the solutions that run faster than the speed of light, $|\gamma|>1$ are always pathological. However, this is not sufficient for avoiding CTCs. We should remember that the solution is constructed by adding fluxes to a four-dimensional Euclidean base space with regions of $(-,-,-,-)$ signature, and it is only the presence of magnetic fluxes that render these solutions regular. Hence, as expected, when $q$ is very small CTC's can reappear.

To facilitate comparison with the quadrupole moments of other black holes, it is natural to introduce the dimensionless multipole ratio \cite{Bena:2020see} 
\begin{equation}
 R\equiv \dfrac{M_2 M }{ J^2}\, .
\end{equation}
The Kerr solution has $R=-1$. 

As one can see from \eqref{M J M2 self dual}, this ratio is positive at small velocities, becomes negative as the bolt velocity increases, and diverges as the bolt velocity approaches the speed of light $\gamma \rightarrow \pm 1 $.

 \subsection{An infinite number of solutions}\label{sec:more solutions}

For clarity of presentation, we restricted the analysis of solutions of equation \eqref{periodicity of phi} to  $\n_1= 1 , 2, 3$. A natural question is whether smooth, horizonless solutions exist for arbitrary integer $\n_1$.

\begin{figure}[h!]
    \centering

\begin{tikzpicture}
        % --- Image de gauche ---
        \node[anchor=south west,inner sep=0] (img) at (0,0)
            {\includegraphics[width=0.45\textwidth]{m-alpha_3.jpeg}};
    \begin{scope}[x={(img.south east)},y={(img.north west)}]
        % Rectangle autour de la zone ˆ zoomer (dans la figure de gauche)
        \draw[violet,thick] (0.15,0.8) rectangle (0.22,0.85);

        % --- Image de droite ---
        \node[anchor=south west,inner sep=0] (zoom) at (1.2,0.16)
            {\includegraphics[width=0.45\textwidth]{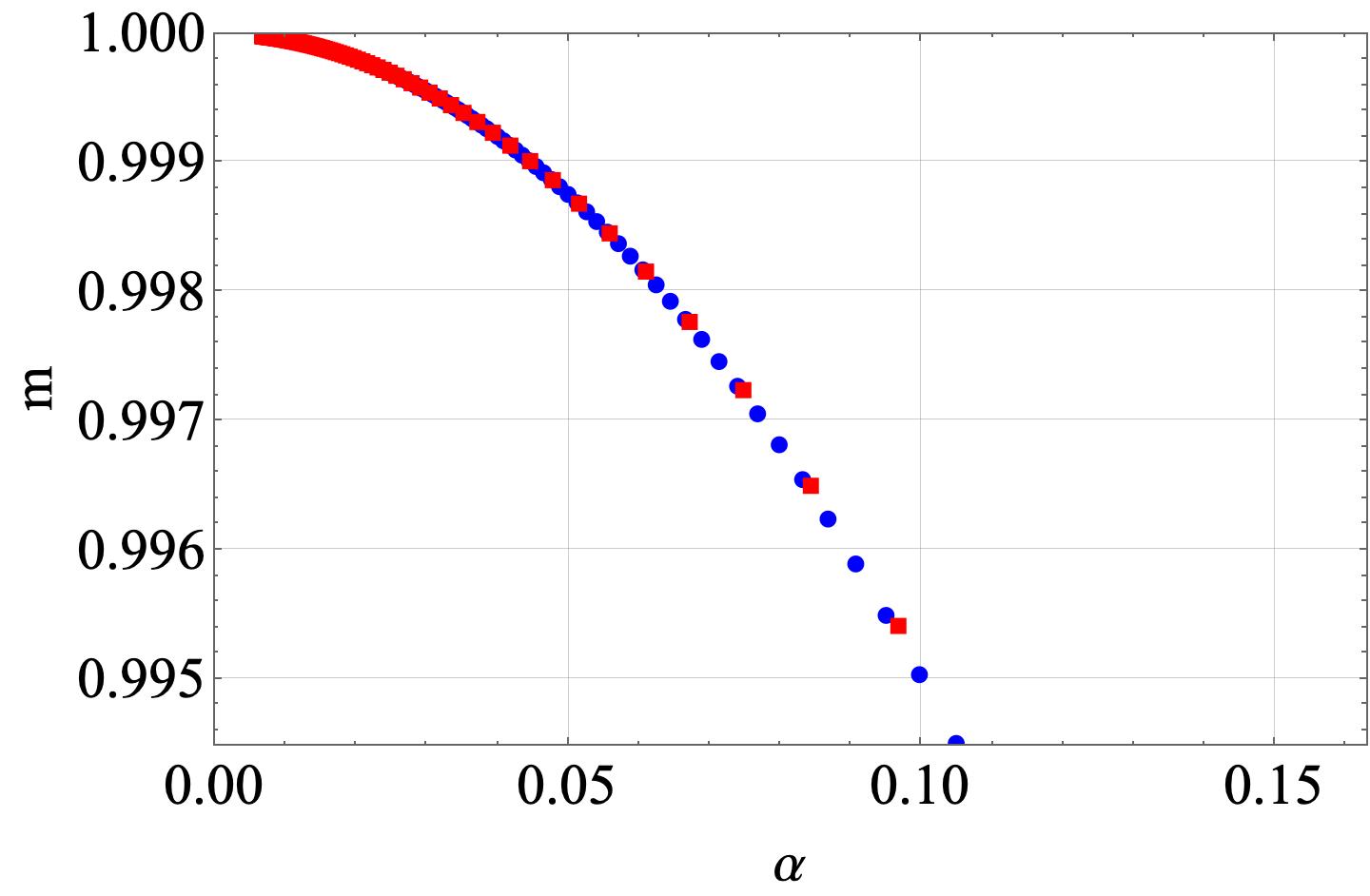}};

        % Connexion entre le rectangle et la deuxime image
        \draw[dashed,thick,violet] (0.22,0.85) -- (1.33,0.84);
        \draw[dashed,thick,violet] (0.22,0.8) -- (1.33,0.28);
        \end{scope}
\end{tikzpicture}
    \caption{Right figure: more $N=1$ solutions $(m,\alpha)$ for values of $\n_1$ from 4 to 100. Square red points are solutions of type 1, and blue  round points are solutions of type 2.}
    \label{fig:zoom on more solutions}

\end{figure}

By applying the same construction and regularity checks used for small $\n_1$, we find that such solutions do indeed exist for arbitrary large $\n_1$. However, eliminating CTCs imposes increasingly stringent constraints on $q$ as $\n_1$ grows. For example, when $\n_1=100$ the range of $q$ where the solutions are regular becomes very narrow: $[0.03456,0.03470]$.

We also find that, in the large-$\n_1$ limit, the solutions with regular asymptotics fall into two classes, satisfying a simple relation between $\alpha$ and $\n_1$ which can be found from expanding \eqref{same-sign}-\eqref{opp-sign} around $m'=1$: 
\begin{equation}
     \left\{
    \begin{array}{ll}
    \alpha_1(\n_1)& \approx \dfrac{2}{ 3 \n_1}\\[1em]
    \alpha_2(\n_1)& \approx  \dfrac{2}{\n_1}
    \end{array}
\right. \, .
\label{fit alpha k}
\end{equation}

It is not hard to see from Figure \ref{fig:zoom on more solutions} that the two classes of solutions exist because a given red curve, $m(\alpha)$, and a given dashed blue parabola (constant $\n_1$) cross twice.
The second branch $\alpha_2$ only becomes visible when $\n_1\geq 10$.

\begin{figure}[h!]
    \centering
    \includegraphics[width=0.5\linewidth]{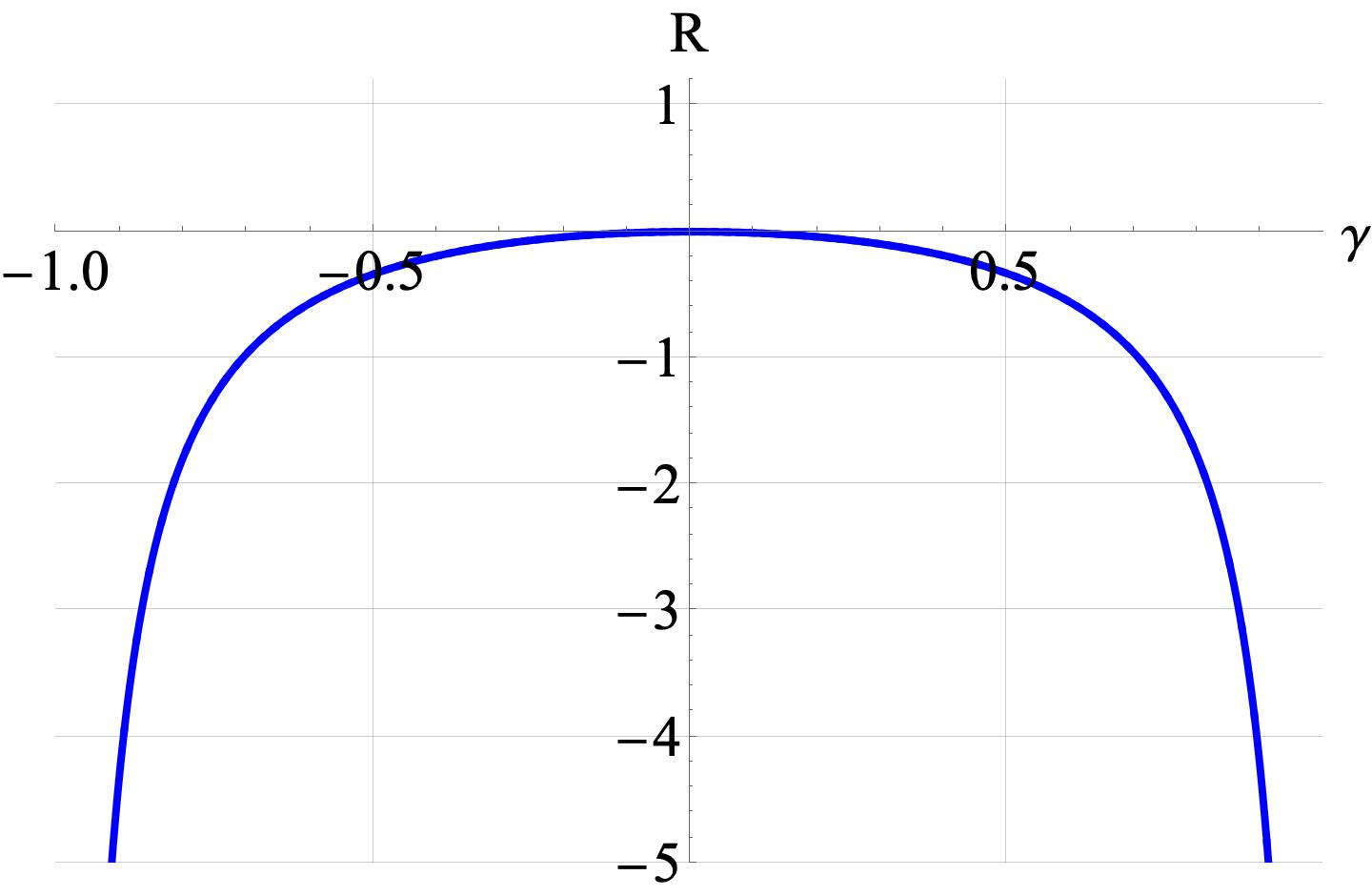}
    \caption{Plot of $R$ as a function of $\gamma$, for $\n_1=100$, corresponding to  $(\alpha_1=6.7.10^{-3},m=0.99998)$. The value of $R$ at $\gamma=0$ is $9.260 .10^{-5}$  and the maximum value of $R$ is slightly higher, $1.008 .10^{-4}$.}
    \label{fig:M2(gamma) for n1=100}
\end{figure}

As one approaches the highly twisted region, in which $\n_1$ increases and $\alpha$ tends to zero, the maximal value of the quadrupole ratio, $R$, and its value at $\gamma=0$ tend also to zero from above. For example, when  $\n_1=100$ one has $-\infty<R<9.26 .10^{-5}$ (see Figure~\ref{fig:M2(gamma) for n1=100}). Therefore, to find solutions with a large positive quadrupole moment one needs to focus on Kerr-Taub-Bolt base spaces of lower $\n_1$.

We can calculate the value of the fluxes, $q_0$ where the solutions do not run, and see analytically that the spin-induced quadrupole moment of non-running solutions is always positive. We also find that the expression for the octopole moment, $M_3$  simplifies drastically and is generically nonzero:
\begin{equation}
     \left\{
    \begin{array}{ll}
    q_0 \equiv \sqrt{6} (N-m) \sqrt{\dfrac{r_+}{2 r_+-m-N}}\\[1em]
    G_4M_2|_{\gamma=0} =\dfrac{\alpha ^2 (N-m) (N+m) \left(16 \sqrt{\alpha ^2+m^2-N^2}+17 m+N\right)}{2 \left((9N-7m) \sqrt{\alpha ^2+m^2-N^2}+8 m(N-m)\right)}\ >0 \\[1em]
    G_4 M_3|_{\gamma=0} = \dfrac{\alpha ^3 (N-m) (N+m) (16 \alpha +\n_1 (17 m+ N)) \left(\n_1(9 m^2-8 m N- N^2)- \alpha (7N-9m)\right)}{\left(8 \n_1 m(N-m)+ \alpha (9 N-7 m)\right)^2}\\[1em]
        \end{array}
\right. \, .
\end{equation}

It is also possible to evaluate the asymptotic behavior of the charges and multipole moments of the $\gamma=0$ solutions in the large-$\n_1$ limit. 
\begin{equation}
     \left\{
    \begin{array}{ll}
    q_0|_{\gamma=0} &\underset{\n_1 \rightarrow\infty}{\sim} \dfrac{\sqrt{3}\ C_i^2}{\n_1}\sqrt{\dfrac{1}{C_i(4-C_i)}}\\[1em]
    G_4M_2|_{\gamma=0} &\underset{\n_1 \rightarrow\infty}{\sim} \dfrac{C_i^2}{4 k^2} \dfrac{18 C_i}{1+2C_i}\ >0\\[1em]
    G_4 M|_{\gamma=0}  & \underset{\n_1 \rightarrow\infty}{\sim} \dfrac{5 C_i+4}{2 (4-C_i)}\ >0\\[1em]
    G_4 J|_{\gamma=0} &\underset{\n_1 \rightarrow\infty}{\sim} \dfrac{6 \sqrt{3} C_i^2}{(4-C_i) \sqrt{(4-C_i)C_i}}\, 
        \end{array}
\right. \, ,
\end{equation}
where $C_1 = \dfrac{2}{3}$ and $C_2 = 2$ for the two classes of solutions in \eqref{fit alpha k}
.

Note that, even if the parameters of the base appear to approach the Taub-NUT values in the limit $p\rightarrow \infty$, the solutions differ drastically. For example, the Taub-NUT solutions, which we will review in the next section, have zero $J$, while the large-$p$ Running-Kerr-Taub-Bolt solutions have finite $J$.

%%%%%
\section{Running-Bolt solution with a Taub-NUT base space}
%%%%%
\label{section5}

The purpose of this section is to show that when $m=\pm N$, the Kerr-Taub-Bolt metric reduces to a Gibbons-Hawking metric, and the running Kerr-Taub-Bolt solutions become supersymmetric fluxed Taub-NUT (TN) solutions in the class of supersymmetric solutions constructed in \cite{Gauntlett:2004wh,Bena:2005ni}. Since $m$ is always positive, there are two classes of TN solutions one can construct, with positive  NUT charge ($N=m$) and with negative NUT charge ($-N=m$). Solutions with positive $N$ only admit self-dual fluxes and solutions with negative $N$ only admit anti-self-dual fluxes. We can see that when $N=m$, self-dual solutions diverge, reflecting the non-normalizability of the self-dual fluxes on a Taub-NUT space with a self-dual complex structure. Similarly, when $-N=m$, the Taub-NUT space changes orientation, and the anti-self-dual solutions diverge.  
 
One can construct $N=m$ or $N=-m$ solutions with good 4D asymptotics both when $\alpha=0$ and when $\alpha\neq 0$. For $\alpha=0$ the change of coordinates from the Kerr-Taub-Bolt coordinates to the Gibbons-Hawking coordinates used commonly to write the Taub-NUT metric is 
\begin{align}
\label{change coordinate TN alpha=0}
     \left\{
    \begin{array}{ll}
        r = 
        %m+r_{TN}= 
        m + 2 \rho_{TN}\\[1em]
        \theta =\theta_{TN}\\[1em]
        \tau = 2 \psi_{TN}
        \end{array}
        \right. \, , 
\end{align}
and the Kerr-Taub-Bolt metric can be written as:
\begin{equation}
\label{eq:metric TN}
\begin{split}
        ds^2_4 &= \frac{r+m}{r-m}\Big(dr^2 + (r-m)^2(d\theta^2+\sin^2 \theta d\phi^2)\Big)+\frac{r-m}{r+m}\left(d\tau +2\beta m \cos \theta d\phi\right)^2 \\
        &= 4 \left( V \Big(d\rho_{TN}^2 + \rho_{TN}^2(d\theta_{TN}^2+\sin^2 \theta_{TN} d\phi^2)\Big)+\frac{1}{V}\left(d\psi_{TN} +\beta  m \cos \theta_{TN}d\phi\right)^2 \right)\ ,
\end{split}
\end{equation}
where $N=\beta m$, with $\beta = \pm 1$ and $\psi_{TN}$ is $4 \pi m$ periodic. For simplicity we will focus on self-dual solutions with $\beta=-1$; one can repeat the analysis for $\beta=1$ straightforwardly.
It is not hard to see that the self-dual Running-Kerr-Taub-Bolt solution becomes a solution with a Gibbons-Hawking base-space determined by 8 harmonic functions\footnote{Since we are interested in a solution with no Wilson lines at infinity we take the constants in the $K_I$ harmonic functions to vanish.}
\begin{equation}
 V = 1+ \frac{m}{\rho_{TN}}~;~~K_I =  \frac{-q_I}{2 \,\rho_{TN}}~;~~ L_I \equiv 1 + \frac{l_I}{\rho_{TN}} ~;~~ M = m_0 + \frac{m_1}{\rho_{TN}}\,.
\end{equation}
The regularity of the Taub-NUT solution at $\rho_{TN}=0$ requires that \cite{Bena:2005va,Berglund:2005vb,Bena:2007kg}
\begin{equation}
l_I=-{1\over 2}C_{IJK}\, {q_J \, q_K \over 2^2~m}~,~~~ m_1 = -{1\over 12}C_{IJK} {q_I q_J q_K \over 2^3 ~m^2}~,~~~m_0 = \frac{4 m^2(q_I+q_J+q_K) - q_I q_J q_K}{16 m^3} \,.  
\end{equation}

%\IB{check factors of 2} \AL{Done}
It is not hard to see that the warp factor and the rotation parameters of the fluxed Taub-NUT solution
\begin{equation}
Z_I = \dfrac{K_J K_K}{V} + L_I~,~~~\mu = \frac{K_I K_J K_K}{V^2}+\frac{1}{2}\frac{K_1 L_1+K_2 L_2+K_3 L_3}{V}+M
\end{equation}
are identical to those of the Running-Kerr-Taub-Bolt solution with $N=-m$ and $\alpha=0$. 
The rotation parameter, $\nu$, vanishes, consistent with the fact that for a single-center fluxed Taub-NUT solution, the four-dimensional rotation parameter  always vanishes. 

Note that the asymptotic value of $\mu$ at infinity does not generically vanish:
\begin{equation}
 \mu \rightarrow \gamma_{TN} = \dfrac{4 m^2 (q_I+q_J+q_K)-q_I q_J q_K}{16 m^3} \,,
\end{equation}
indicating that generic fluxed supersymmetric Taub-NUT solutions are running. Both the angular momentum and the higher mass multipole moments of these solutions vanish, as expected. 

It is a bit harder to check that the $N=-m$ solutions with non-zero $\alpha$ are also fluxed Taub-NUT solutions. They appear to rotate but, if one evaluates their angular momentum, their quadrupole moment \eqref{M J M2 self dual}, and their higher multipole moments, these vanish identically, despite the presence of a nontrivial rotation parameter, $\alpha$, and of a nonzero four-dimensional rotation parameter,  $\nu$.

The change of coordinate to bring the $m=-N$ solutions with $\alpha\neq 0$ to Taub-NUT form is slightly more complicated  \cite{HPS}:
\begin{align}
\label{change coordinate TN alpha != 0}
     \left\{
    \begin{array}{ll}
        r -\alpha \cos \theta= %m+r_{TN} = 
        m+2\rho_{TN} \\[1em]
        (r-m) \cos\theta %=r_{TN}\cos\theta_{TN}+\alpha
    =2\rho_{TN}\cos\theta_{TN}+\alpha\\[1em]
        \tau =2 \psi_{TN} \\
        \end{array}
        \right. \, ,
\end{align}
and the $m=-N$ Euclidean-Kerr-Taub-Bolt metric written in these coordinates is:
\begin{equation}
    ds^2_4 =4\left(\!V \Big(d\rho_{TN}^2 + \rho_{TN}^2(d\theta_{TN}^2+\sin^2 \theta_{TN} d\phi^2)\Big)+\frac{1}{V}\left(d\psi_{TN}\! +\!\left(\dfrac{m^2\alpha/2}{m^2-\alpha^2}-  m \cos \theta_{TN}\!\right)\! d\phi \!\right)^2 \right)\!.
\end{equation}

\vspace*{-.2cm}
\section{Conclusions}
\label{Conclusions}
In this paper, we identified the rotating Euclidean-Kerr-Taub-Bolt solutions with asymptotically-$\mathbb{R}^{3,1}$ four-dimensional asymptotics. This was achieved by taking into account the nontrivial periodicity constraint between the Kaluza-Klein angle and the four-dimensional angle of rotation. This constraint enabled us to obtain a discrete set of allowed Euclidean-Kerr-Taub-Bolt base spaces that give rise to asymptotically-$\mathbb{R}^{3,1}$ solutions, parameterized by an , $\n_1$, and by the Euclidean Nut charge, $N$.  

For a given Euclidean-Kerr-Taub-Bolt base space, the full solutions are completely determined by the value of the three types of self-dual two-form fluxes on the bolt, proportional to $q_1, q_2$ and $q_3$.
We analyzed the regularity of these solutions for equal fluxes, and found that they are regular only in a certain flux interval, where the fluxes are not too small, and the velocity of the Running Bolt is less than the speed of light. 

We computed the spin-induced quadrupole moment of these solutions and found that, for a bolt running at low speeds, the multipole moment is always positive. As explained in the Introduction, this very counterintuitive behavior comes from the String-Theory ingredients of these solutions, which can have unusual properties when regarded from a purely four-dimensional perspective.

There are several open questions that our results raise:

First, what is the relation between our solutions and the spinning solutions  constructed in parallel to our work by Heidmann, Pani and Santos \cite{HPS} by applying Ehlers transformations on the ``Euclidean-Kerr-Taub-Bolt times time" solutions. These solutions are constructed by dualities, which does not allow them to leave the ``duality orbit'' of the original solution\footnote{For example, if the original solution does not have charges and angular momenta  corresponding to a black hole with a large horizon area, nor will the solutions obtained by dualities have these charges. }. In contrast, the Running-Kerr-Taub-Bolt solutions belong to different duality orbits than the ``Kerr-Taub-Bolt times time'' solution. Hence, we expect that applying the dualities of \cite{HPS} on our solutions will allow to swipe these new duality orbits and will produce new classes of solutions. 

It is interesting to note that the solutions of \cite{HPS} can also have positive $M_2$. As we explained above, this may be caused by the presence of ingredients that, from a four-dimensional perspective, have negative mass and negative charge. Remarkably, the solutions of \cite{HPS} are obtained by dualizing ``Kerr-Taub-Bolt times time'' solutions that have regions of $(-,-,-,-)$ signature, which indicates the presence of such ingredients.

The second important question is whether the duality orbit of the Running-Kerr-Taub-Bolt solutions can get inside the charge and angular momentum regime in which nonextremal black holes with a large horizon area exist. This determination is rather subtle, and is left to future work. The best method for determining this \cite{Chakraborty:2025ger} seems to be to use the Chow-Comp\`ere method \cite{Chow:2014cca}  to construct directly the nonextremal black holes with the charges we want. However, this is made complicated by the fact that our solutions are running, which complicates the Chow-Comp\`ere construction and makes the definition of charges subtle. 

It is also well possible that for spinning single-bubble solutions, both Running-Bolts and the solutions of \cite{HPS}, there is an obstruction to being in the black-hole orbit. A similar problem appears to plague solutions constructed using the Bossard-Katmadas method \cite{Bossard:2014yta,Bossard:2014ola,Bena:2015drs,Bena:2016dbw,Bossard:2017vii}.

It would be very important to try to construct generic spinning multi-bubble solutions with black-hole charges (as done for non-spinning black holes in 
\cite{Bah:2022pdn,
Heidmann:2021cms,
Bah:2021owp,
Bah:2021rki,
Heidmann:2022zyd,
Bah:2022yji,
Bah:2023ows,
Heidmann:2023kry} 
and see if one can access the black-hole range and if the unusual spin-induced quadrupole moment we find is a generic feature of such solutions.

It would be very interesting to determine if positive spin-induced quadrupole moments are ubiquitous in Bossard-Katmadas solutions with four-dimensional asymptotics, and also in Bobev-Ruef solitons \cite{Bobev:2009kn} obtained as floating-brane solutions on an Euclidean Einstein-Maxwell base space \cite{Bena:2009fi}.

\noindent{\bf Acknowledgments:} We would like to thank Pierre Heidmann for multiple insights and suggestions. We would also like to thank Guillaume Bossard, Rapha\"el Dulac, Paolo Pani, Jorge Santos, Denis Vion and Nick Warner for interesting discussions.
The work of IB  was supported in part by the ERC Grant 787320 - QBH Structure.

\begin{adjustwidth}{-1mm}{-1mm} % to adjust the L and R margins 

\bibliographystyle{utphys}      

\bibliography{biblio}  
\end{adjustwidth}
\end{document}